\documentclass{nature}
\usepackage{graphicx}

\usepackage[left=1.50cm, right=1.50cm, top=1.00cm, bottom=1.50cm]{geometry}
\usepackage{float}
\usepackage{amsmath}
\usepackage{amssymb}
\usepackage{ulem}
\usepackage{xcolor}
\usepackage[export]{adjustbox}
\usepackage{url}

\usepackage[switch,pagewise]{lineno}

\DeclareRobustCommand{\ion}[2]{\textup{#1\,\textsc{\lowercase{#2}}}}
\newcommand*\element[1][]{%
  \def\aa@element@tr{#1}%
  \aa@element
}

\title{Measuring the ionisation fraction in a jet from a massive protostar}

\author{R.~Fedriani$^{1,2*}$, A.~Caratti o Garatti$^{1,2}$, S.~J.~D.~Purser$^1$, A.~Sanna$^3$,  J.~C.~Tan$^{4,5}$, R.~Garcia-Lopez$^1$, T.~P.~Ray$^1$, D.~Coffey$^{2,1}$, B.~Stecklum$^6$ \& M.~Hoare$^7$}

\date{\\$^1$Dublin Institute for Advanced Studies, School of Cosmic Physics, Astronomy \& Astrophysics Section, 31 Fitzwilliam Place, Dublin 2, Ireland. $^2$University College Dublin, School of Physics, Belfield, Dublin 4, Ireland. $^3$Max-Planck-Institut f\"ur Radioastronomie, Auf dem H\"ugel 69, 53121 Bonn, Germany. $^4$Dept. of Space, Earth \& Environment, Chalmers University of Technology, SE-412 93 Gothenburg, Sweden. $^5$Dept. of Astronomy, University of Virginia, 530 McCormick Road, Charlottesville, VA 22904-4325, USA. $^6$Th\"uringer Landessternwarte Tautenburg, Sternwarte 5, 07778 Tautenburg, Germany. $^7$School of Physics and Astronomy, University of Leeds, Leeds LS2 9JT, UK.}

\begin{document}

\maketitle


\begin{abstract}
\noindent\textbf{Abstract}\\
It is important to determine if massive stars form via disc accretion, like their low-mass counterparts. Theory and observation indicate that protostellar jets are a natural consequence of accretion discs and are likely to be crucial for removing angular momentum during the collapse. However, massive protostars are typically rarer, more distant and more dust enshrouded, making observational studies of their jets more challenging. A fundamental question is whether the degree of ionisation in jets is similar across the mass spectrum. Here we determine an ionisation fraction of $\sim5-12\%$ in the jet from the massive protostar G35.20-0.74N, based on spatially coincident infrared and radio emission. This is similar to the values found in jets from lower-mass young stars, implying a unified mechanism of shock ionisation applies in jets across most of the protostellar mass spectrum, up to at least $\sim10$ solar masses.
\end{abstract}


\noindent\textbf{Introduction}\\
Although massive stars are rare, they play a fundamental role in the Universe, synthesising most of the chemical elements and providing a major feedback into the molecular clouds where stars are born (see ref.\cite{tan2014} and references therein for a review). However, massive star formation and evolution are still a matter of debate. An excellent tool to investigate massive star formation is provided by protostellar jets. The detection of several jets driven by massive protostars\cite{caratti2015,fedriani2018,sanna2015,sanna2019} and the discovery of dusty molecular discs around high-mass young stellar objects (HMYSOs, $M_{*}>8\,M_\odot, L_\mathrm{bol}>5\times10^3L_\odot$) through near-infrared (NIR) interferometry\cite{kraus2010}, strongly support the idea that HMYSOs form in a similar way to low-mass young stars\cite{kraus2010,ilee2013,beltran2016,caratti2017nature,cesaroni2018}. Collimated jets associated with discs of few 100\,au have also been observed at mm wavelengths, supporting this scenario\cite{patel2005,girart2018}. The jets themselves are thought to be launched centrifugally along magnetic field lines\cite{pudritz2007} and thus the jet's ionisation fraction ($\chi_e$) is a key parameter as determines the strength of the coupling of the magnetic field to the ionised gas. Often, when irradiated by a nearby OB star, the jet is considered to be fully ionised or alternatively a somewhat arbitrary value is used\cite{bally2006a,purser2016,mcleod2018}. In a number of cases for low-mass young stars, $\chi_e$ has been measured using emission line diagnostics in the optical/NIR regime\cite{podio2006}. Determining $\chi_e$ is important if we are to correctly deduce such parameters as the total (neutral plus ionised) mass-loss rate in the jet  $(\dot{M}_\mathrm{ejec})$, which, in turn, is often compared with the mass-accretion rate $(\dot{M}_\mathrm{acc})$. Since dynamical quantities are fundamental inputs in massive star formation models, a good constraint on the ionisation fraction along massive jets is critical. The radio continuum emission from protostellar jets is generally interpreted as thermal bremsstrahlung, and, since the radio emission does not suffer from extinction due to dust, it is an excellent way of observing the ionised component\cite{purser2016,anglada2018,sanna2018}. However, no velocity information can be inferred from these observations (unless we rely on proper motion studies and we know the jet geometry) nor can the ionisation fraction be directly derived. On the other hand, NIR spectroscopy directly provides us with physical parameters and dynamical information for the (molecular and atomic) jet through the study of emission lines\cite{caratti2015,fedriani2018}. Therefore, radio and NIR observations of both line and continuum emission allow for a complementary analysis of protostellar jets.

\begin{figure}[ht!]
\includegraphics[width=0.5\textwidth]{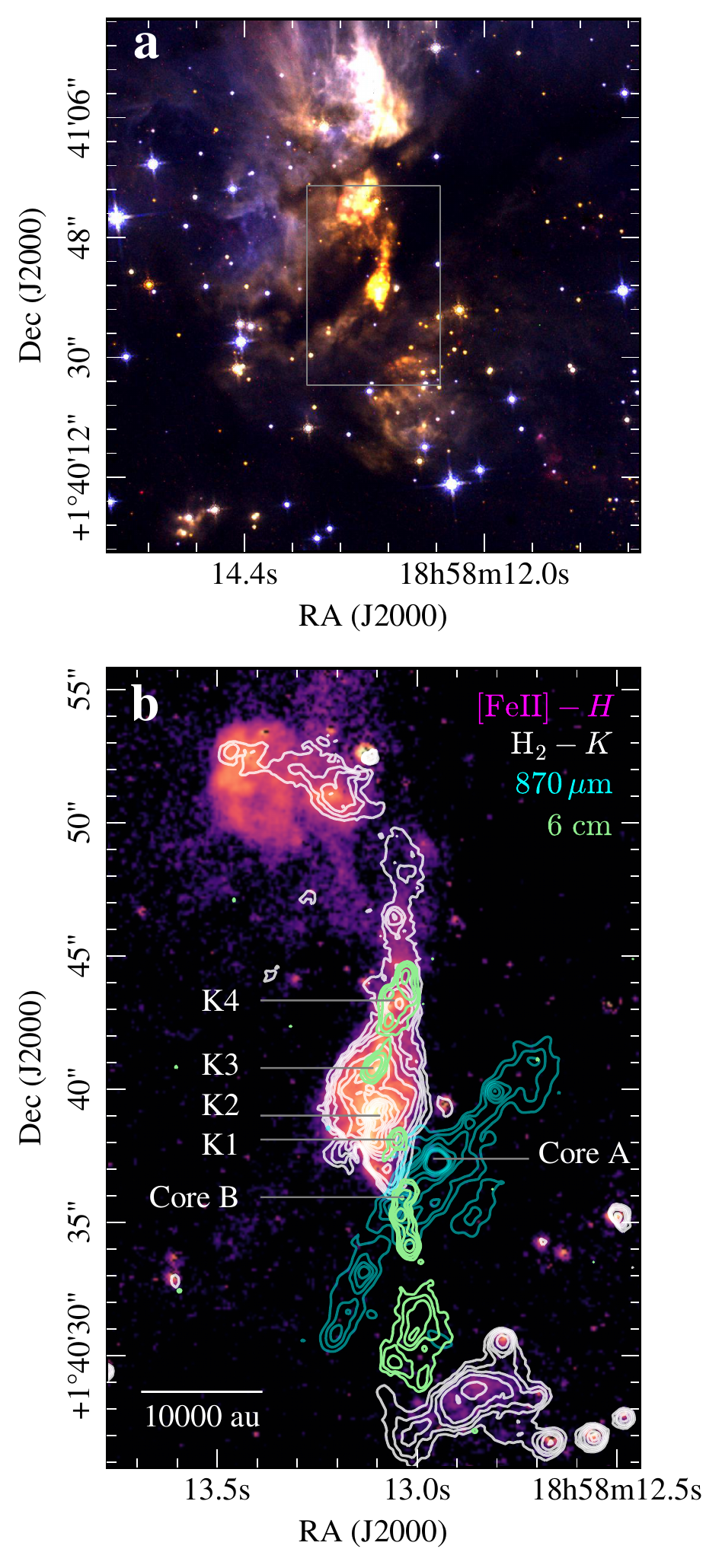}
\caption{{\footnotesize G35.2-0.74N star formation complex. \textbf{a} Three colour composite HST image of the innermost region of the star-forming site G35.20-0.74N, revealing the jet from the massive protostar. Red channel is the [\ion{Fe}{ii}] 1.644\,$\mu$m narrow filter (F164N), green channel is the $H$ 1.600\,$\mu$m wide filter (F160W), and blue channel is the $J$ 1.100\,$\mu$m wide filter (F110W). \textbf{b} Composition of the HST/WFC3 (violet) image, VLT/ISAAC (white contours), ALMA (cyan), and VLA (green) datasets. In both panels, north is up and east is left. The position of the main cores in the region as well as the jet knots are labelled. The field of view of panel b is indicated with a rectangle in panel a.}}
\label{fig:intro}
\end{figure}

The well-known high-mass star-forming region G35.20-0.74 is located at $2.2\,$kpc\cite{zhang2009} in the tail of the Aquila constellation. G35.20-0.74N (hereafter G35.2N) is a main formation site of B-type stars, has a bolometric luminosity of $3\times10^4\,L_\odot$ and hosts two main cores, core A and core B. Both cores display discs in Keplerian rotation\cite{sanchez-monge2014,beltran2016G35N}. Core B is a binary system which consists of two B-type protostars with masses of $11$ and $6\,M_\odot$\cite{beltran2016G35N}, sources 8a and 8b, respectively. The rotation axis of the disc, i.e. the jet axis, of source 8a has an inclination angle of $i\sim19\pm1^\circ{}$ with respect to the plane of the sky\cite{sanchez-monge2013}. Perpendicular to this disc a radio jet close to the central engine has been detected\cite{gibb2003,beltran2016G35N} as well as a molecular hydrogen (H$_2$) outflow\cite{lee2012,caratti2015}. Source 8a is thought to drive one of two parsec-scale bipolar outflows\cite{caratti2015,beltran2016G35N} in this region with an initial north (blue-shifted lobe) south (red-shifted lobe) orientation (see Fig.~3 of ref.\cite{caratti2015} for a complete view of both parsec-scale jets). Here, we focus on the protostellar jet driven by source 8a (see Fig.~\ref{fig:intro}).

Here, we show a unique example of spatially coincident NIR and radio jet emission. We use multi-wavelength observations, i.e. data from the Hubble Space Telescope (HST), Karl G. Jansky Very Large Array (VLA), and Very Large Telescope (VLT), of the outflow from G35.20-0.74N to determine the ionisation fraction, $\chi_e$, in a jet from a massive young star. The values found, $\sim5-12\%$, are similar to those found for solar mass young stars. Our observations confirm that the ionising mechanism giving rise to the radio emission originates from shocks seen in the NIR jet\cite{anglada2018}.


\noindent\textbf{Results}\\
\textbf{NIR imaging and spectroscopy in G35.2N}\\
We performed high-resolution NIR imaging and long-slit spectroscopy of G35.2N using the infrared spectrometer and array camera (ISAAC) at the VLT. We performed imaging to study in detail the NIR jet and spectroscopy to measure its kinematics and derive its physical and dynamic properties. We combine our observations with VLA and HST data to form the most complete view of a jet close to its source to date (see Fig.~\ref{fig:intro}, panel b). We observed the atomic jet in the form of [\ion{Fe}{ii}] and Br$\gamma$ (hydrogen recombination line) in the NIR as well as the ionised jet in the form of \ion{H}{ii} (ionised hydrogen) in the radio. G35.2N, thus, represents a unique example because both the NIR and radio jet are visible and we see that both atomic and ionised emission are spatially coincident. Therefore, we can combine the information from both regimes and infer the ionisation fraction in a HMYSO jet.

Figure~\ref{fig:intro}, panel a shows an HST red-green-blue (RGB) image ([\ion{Fe}{ii}], $H$, $J$, respectively) of the G35.2N central region obtained with the Wide Field Camera 3 (WFC3). Panel b shows a composite image of the inner few $10\,000\,$au from the central star revealing the various components of the jet emission of G35.2N. The HST [\ion{Fe}{ii}] continuum-subtracted image (F164N-F160W, which corresponds to the [\ion{Fe}{ii}] emission at $1.644\,\mu$m) is shown in violet, and it is indicative of jet shocked material\cite{nisini2002,garcia-lopez2008,fedriani2018}. White contours are ISAAC H$_2-K$ emission (corresponding to the H$_2$ emission at $2.121\,\mu$m), which mostly delineate the outflow cavity walls\cite{caratti2015}. Green contours are VLA C-band emission at $6\,$cm (5.8\,GHz), that comes from the ionised radio jet. Notably, the radio data, not affected by extinction, trace both the blue- and red-shifted lobes, whereas the NIR emission mostly comes from the blue-shifted jet. The position of the radio source driving the jet is labelled as Core B and the main knots of study are labelled as K1, K2, K3, and K4. The cyan contours show the Atacama Large Millimeter Array (ALMA) $870\,\mu$m emission from dust, which reveals the locii where dust and gas condense into stars\cite{sanchez-monge2013}.

\begin{figure*}[ht!]
\centering
\includegraphics[width=1.0\textwidth]{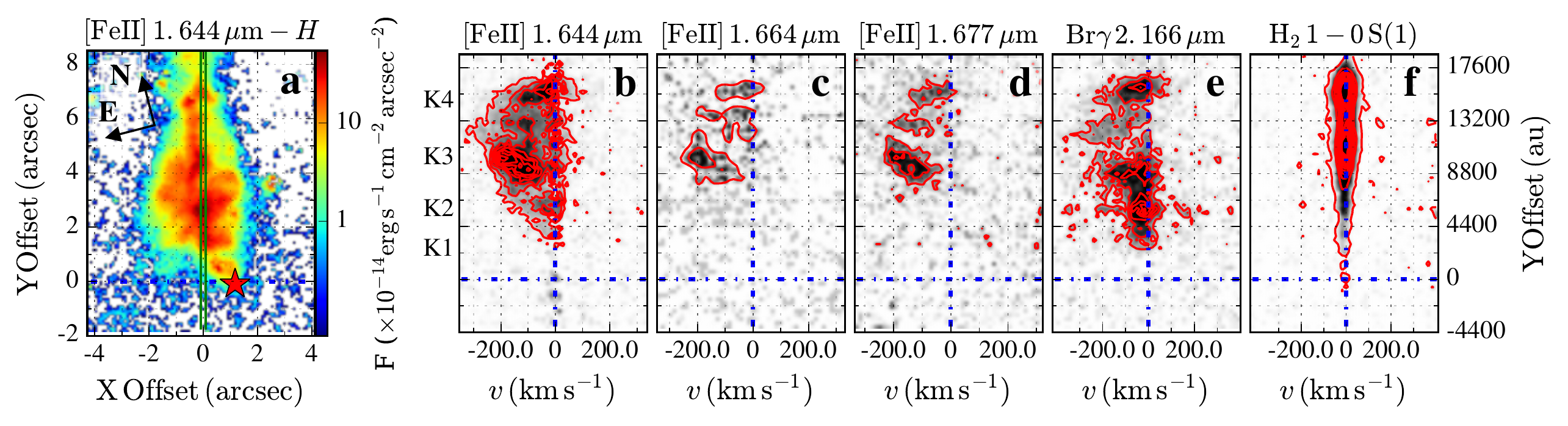}
\caption{{\footnotesize Position-velocity diagram of the G35.2N protostellar jet. \textbf{a} HST [\ion{Fe}{ii}]$-H$ image oriented in the slit position (green lines, PA = $-14.7^\circ{}$), north and east are recorded in the top-left corner of the figure, the central source is indicated with a red star, and the flux values are indicated in the colour bar to the right. \textbf{b} ISAAC spectral image of the [\ion{Fe}{ii}] line at 1.644\,$\mu$m. \textbf{c} ISAAC spectral image of the [\ion{Fe}{ii}] line at 1.664\,$\mu$m. \textbf{d} ISAAC spectral image of the [\ion{Fe}{ii}] line at 1.677\,$\mu$m. \textbf{e} ISAAC spectral image of the Br$\gamma$ line at 2.166\,$\mu$m. \textbf{f} ISAAC spectral image of the H$_2$ line at 2.121\,$\mu$m (1--0 S(1)). The red contours represent the intensity in the spectral images and are from $3\sigma$ to $40\sigma$ in steps of $3\sigma$. The horizontal blue dashed lines indicate the position $0''$ in the Y Offset while the vertical blue dashed lines indicate the position of $0\mathrm{\,km\,s^{-1}}$ in panels b to f.}} 
\label{fig:PVD}
\end{figure*}

The spectral images of the protostellar jet close to the source are shown in the position-velocity (PV) diagram (Fig.~\ref{fig:PVD}). Our spectroscopic slit, at a position angle (PA) of $-14.7^\circ{}$, encompasses knots K2, K3, and K4. The velocities are measured with respect to the Local Standard of Rest (LSR) and subsequently corrected by the velocity of the parent cloud ($v_\mathrm{LSR}\sim33\mathrm{\,km\,s^{-1}}$, ref. \cite{sanchez-monge2013,sanchez-monge2014}). H$_2$ emission, namely the $1-0$\,S(1) transition at $2.121\,\mu$m and the $2-1$\,S(2) transition at $2.154\,\mu$m, is observed towards the outflow (see Fig.~\ref{fig:PVD}, panel f for the spectral image of the $1-0$\,S(1) line). The radial velocity ($v_\mathrm{rad}$) of the H$_2$ lines peaks at red-shifted velocities of $\approx 5\mathrm{\,km\,s^{-1}}$, corresponding to a total velocity of $v_\mathrm{tot}\approx 15\mathrm{\,km\,s^{-1}}$ ($v_\mathrm{tot}=v_\mathrm{rad}/\sin i$, where $i$ is the inclination of the jet axis with respect to the sky), which might be an indication of slow oblique shocks against the cavity walls. [\ion{Fe}{ii}] emission is also detected along the jet (see Fig.\,\ref{fig:PVD}, panels b, c, and d). Blue-shifted radial velocities range from few $\mathrm{\,km\,s^{-1}}$ to $-200\mathrm{\,km\,s^{-1}}$, which correspond to a total velocity of $v_\mathrm{tot}\approx -600\mathrm{\,km\,s^{-1}}$. We also observe Br$\gamma$ emission at $2.166\mu$m (hydrogen recombination line transition from $n_{up}=7$ to $n_{low}=4$). This emission is observed to be extended up to $\sim8''$ from the central source (i.e. $\sim17\,600$\,au at a distance of $2.2\,$kpc; see Fig.~\ref{fig:PVD}, panel e). This emission covers the same velocity range of the [\ion{Fe}{ii}] emission, consistent with proper motions given by ref.\cite{beltran2016G35N}. Emission from [\ion{Fe}{ii}] and Br$\gamma$ is, thus, indicative of shocked material\cite{garcia-lopez2008,fedriani2018}. This atomic emission is spatially coincident with the radio jet emission towards the north (blue-shifted lobe) where the visual extinction ($A_\mathrm{V}\approx 25$\,mag\cite{fuller2001}) allows the detection of NIR lines. Conversely, towards the south (red-shifted lobe), the visual extinction reaches up to $\sim41$\,mag (increasing up to $\sim170$\,mag towards the driving source and its immediate surroundings) \cite{fuller2001} hindering any NIR jet-detection, although the red-shifted radio jet is still visible (Fig.~\ref{fig:intro}, panel b). To investigate the possibility that some of the Br$\gamma$ emission is due to scattered light, we have thoroughly checked the PV diagram of Figure~\ref{fig:PVD}. Br$\gamma$ emission close to $0\mathrm{\,km\,s^{-1}}$ is likely scattered emission from the central source. But, there is also material moving radially at more than $-150\mathrm{\,km\,s^{-1}}$ (which corresponds to $v_\mathrm{tot}\geq-500\mathrm{\,km\,s^{-1}}$). Moreover, the full width at zero intensity (FWZI), which gives an estimate of the jet shock velocity\cite{hartigan1987}, is larger than $500\mathrm{\,km\,s^{-1}}$. This value is consistent with the total velocities derived above, supporting the jet geometry. Finally, the [\ion{Fe}{ii}] emission spatially coincides with that from Br$\gamma$ and their velocities are similar. This evidence suggests that the Br$\gamma$ line is likely tracing shocked material as [\ion{Fe}{ii}] does. This is also confirmed by the fact that all Br$\gamma$ emission is blue-shifted which is consistent with the lobe being directed towards us. Thus, the Br$\gamma$ emission likely comes from a combination of scattered light and from directly visible shocked material. Notably, the radio continuum, which traces ionised gas, emits co-spatially  with the atomic emitting region implying that the ejected material is partially ionised.

\begin{figure}[ht!]
\centering\includegraphics[width=0.5\textwidth]{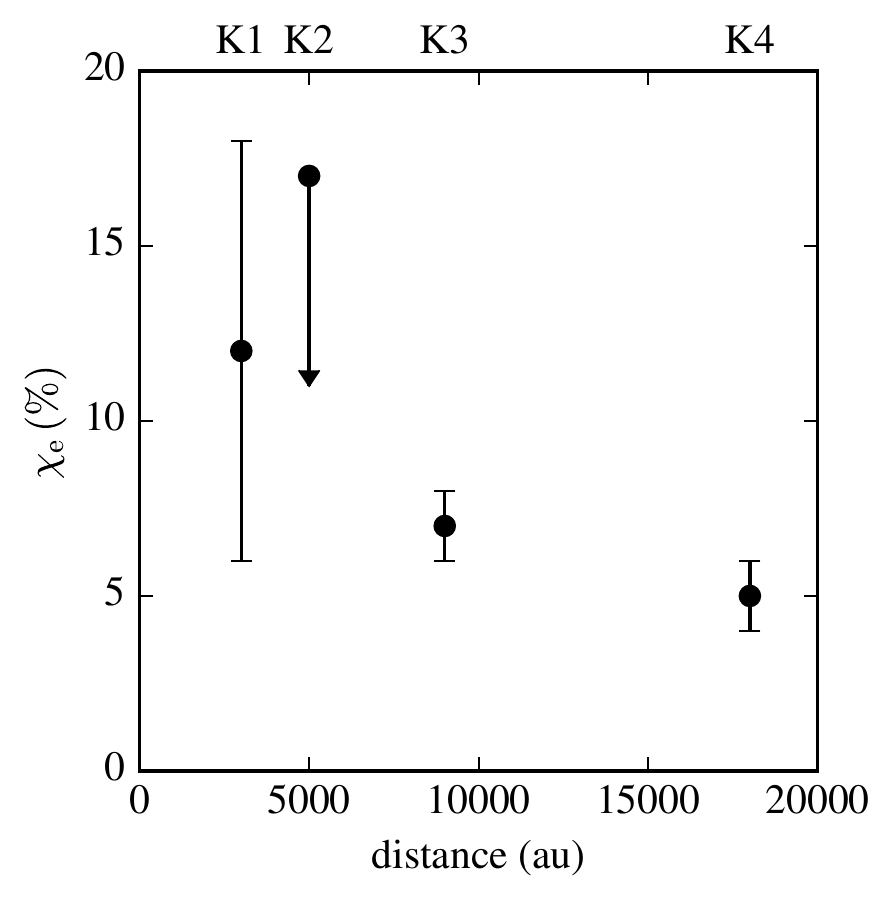}
\caption{Ionisation fraction versus distance. Evolution of the ionisation fraction along the jet with distance to the source. The source of uncertainties is coming from the error propagation calculated from the ionisation fraction equation and the errors are $6\%$ (K1), $1\%$ (K3), and $1\%$ (K4), K2 being an upper limit.}
\label{fig:ionisation_fraction_vs_distance}
\end{figure}

\noindent\textbf{Determination of ionisation fraction and dynamic properties}\\
We can determine the degree of ionisation of the jet from the massive protostar by directly comparing the total number density $(n_\mathrm{tot})$ with the electron number density $(n_\mathrm{e})$. The ionisation fraction is defined as $\chi_\mathrm{e}=n_\mathrm{e}/n_\mathrm{tot}$. On the one hand, the $n_\mathrm{tot}$ can be derived from the properties of the [\ion{Fe}{ii}] emission line at $1.644\,\mu$m (among the brightest transitions in the NIR). The product of the total number density times the volume emitting region ($n_\mathrm{tot}V$) can be written as the ratio between the observed luminosity of the line and the emissivity per particle for the line calculated theoretically \cite{podio2006}, i.e., $n_\mathrm{tot}V=L_\mathrm{[\ion{Fe}{ii}]}/\epsilon_\mathrm{[\ion{Fe}{ii}]}$. The radius and the length of the jet are resolved in our HST and ISAAC observations, therefore the [\ion{Fe}{ii}] emitting volume is known. In this way we can have an estimate of the total number density, given by $n_\mathrm{tot} = L_\mathrm{[\ion{Fe}{ii}]}/\epsilon_\mathrm{[\ion{Fe}{ii}]}/V$ (see Methods for more details). On the other hand, the $n_\mathrm{e}$ can be derived from the ratio of the [\ion{Fe}{ii}] lines\cite{nisini2002,takami2006} or from the properties of the radio emission\cite{mezger1967} (see Methods). Interestingly, both independent methods provide us with the same electron density estimates within the errors. As the radio regime here provides us with smaller errors, we then adopt these electron density estimates. We are able to give a direct measurement of the ionisation fraction in a HMYSO jet for knots K1 ($12\pm6\%$), K3 ($7\pm1\%$), and K4 ($5\pm1\%$). Additionally, in the case of K2 ($<17\%$) we are able to set an upper limit due to the non-detection of radio emission. Figure~\ref{fig:ionisation_fraction_vs_distance} shows a plot of the variation of the ionisation fraction with distance where there is not a strong indication that the $\chi_\mathrm{e}$ changes with distance as the inferred values are the same within the error bars. It is worth noting that the uncertainty in the knot sizes does not significantly affect the ionisation fraction measurement of knots K3 and K4, whereas in the case of K1, the uncertainty in the knot size represents half of the error of the ionisation fraction. Finally, we would like to stress that the determination of the ionisation fraction rests under a few reasonable assumptions, namely $i)$ all Fe is ionised, $ii)$ the Fe abundance is solar, and $iii)$ there is no Fe dust depletion (see Methods for a detailed discussion of these assumptions).

Now, using the mass of each knot, its velocity, and its length, we can also determine the total mass-loss rate using the [\ion{Fe}{ii}] emission line at $1.644\,\mu$m (see Methods). We obtain a mass-loss rate of $4.7\pm0.3\times10^{-5}\,M_\odot\mathrm{\,yr^{-1}}$ and $4.9\pm0.4\times10^{-5}\,M_\odot\mathrm{\,yr^{-1}}$ for knots K3 and K4, respectively. In the case of knot K1 there is no velocity information, because our spectroscopic slit does not encompass it, and in the case of knot K2 just an upper limit on $n_\mathrm{e}$ could be inferred, therefore no dynamic properties were derived. The momentum rate is $0.018\pm0.001\,M_\odot\mathrm{\,yr^{-1}\,km\,s^{-1}}$ and $0.011\pm0.001\,M_\odot\mathrm{\,yr^{-1}\,km\,s^{-1}}$ for knots K3 and K4, respectively.

The total mass-loss rate can also be calculated using the properties of radio emission for these knots in combination with the ionisation fraction derived above. In an analogous way, we combine the size of the radio knots, the electron density, the tangential velocity from our spectroscopic observations (i.e. $v_\mathrm{tan}=v_\mathrm{rad}/\tan(i)$), and the ionisation fraction, to calculate the total mass-loss rate for knots K3 and K4 (see Methods). It results in $\dot{M}_\mathrm{ejec}=1.6\pm1.0\times10^{-5}\,M_\odot\mathrm{\,yr^{-1}}$ for knot K3 and $\dot{M}_\mathrm{ejec}=2.0\pm1.0\times10^{-5}\,M_\odot\mathrm{\,yr^{-1}}$ for knot K4. Remarkably, these values match very well, within a factor of 2 to 4, with those obtained in the NIR regime. Such a small variation is caused by the slight difference in the size of the NIR and radio emitting regions, with the latter being smaller. This is not unexpected as the jet is supposed to have an onion-like structure (see e.g., ref.~\cite{machida2014,agra-amboage2014}), and the more ionised component is expected to be confined to the innermost region (i.e., closest to the jet axis). The ionised mass-loss rate $(\dot{M}_\mathrm{ionised})$ of the considered radio knots is $\sim10^{-6}\,M_\odot\mathrm{\,yr^{-1}}$ being a factor of $\sim10$ smaller than what we get in the NIR, as expected from the ionisation fraction values obtained here. This is in excellent agreement with the ionised mass-loss rate calculated on source (core B) following Reynolds' formulation\cite{reynolds1986}, which provides a value of $1.8\pm0.3\times10^{-6}\,M_\odot\mathrm{\,yr^{-1}}$, consistent with ref.~\cite{beltran2016G35N} (see Methods).

The mass-accretion rate can also be estimated assuming that the ratio $\dot{M}_\mathrm{ejec}/\dot{M}_\mathrm{acc}$ is between $\sim10\%$ and $\sim30\%$ \cite{cabrit2007,antoniucci2008}. Considering a mass-loss rate of $\sim4-6\times10^{-5}\,M_\odot\mathrm{\,yr^{-1}}$, we obtain a mass-accretion rate of $1.3\times10^{-4}\lesssim\dot{M}_\mathrm{acc}\lesssim6\times10^{-4}\,M_\odot\mathrm{\,yr^{-1}}$ consistent with ref.~\cite{zhang2013,debuizer2017}. A typical jet phase timescale in HMYSOs is $\sim2\times10^4$\,yr\cite{guzman2012,fedriani2018}, which gives a lower limit of the age of the star, and, in turn, an estimate of the mass of the central source being $2.6\lesssim M_*\lesssim12\,M_\odot$, which agrees with previous estimates using other methods\cite{zhang2013,sanchez-monge2013,beltran2016G35N}.


\noindent\textbf{Discussion}\\
In the case of jets from low-mass young stars, the ionisation fraction is typically found to be around 10\% or less\cite{hartigan1994,garcia-lopez2008}. Recent studies of two jets from HMYSOs, tentatively derive similar values in an indirect manner and under different assumptions than used here ($\sim10\%$ in IRAS13481-6124\cite{fedriani2018}, and $\lesssim14\%$ in S255IR NIRS3\cite{cesaroni2018}). In the case of IRAS13481-6124, the authors assume that the mass-loss rate calculated along the parsec-scale jet, traced by the H$_2$, has remained roughly constant for the formation history of the star and, thus, should be similar very close to the star where the actual mass-loss rate could not be calculated. Then, they compare this value with the ionised mass-loss rate inferred from the radio regime very close to the star. With this comparison of mass-loss rates, they give a rough estimate of the ionisation fraction. In the case of S255 NIRS3, the authors estimate the mass-accretion rate from the NIR regime and assume that at least $10\%$ of the accreted material should be ejected. This provides an upper limit to the ionisation fraction by comparing to the ionised mass-loss rate calculated from the radio continuum emission. In any case, these previous estimates were made without a direct evidence of spatially coincident atomic and ionised jet emission. In any event, our result is consistent with these indirect findings, i.e., that the ionisation fraction is low and similar to what is found in the low-mass regime. Additional studies of this kind however are warranted to determine whether this is the norm amongst outflows from HMYSOs. 

In the regions of the jet of G35.2N probed by our observations -i.e. knots K1, K2, K3 and K4- we see a similar ionisation degree as found in low-mass YSOs, suggesting a mechanism of shock ionisation, rather than photo-ionisation, for setting the ionisation degree. However, towards the later stages of massive star formation the protostar is expected to contract close to a zero age main sequence (ZAMS) structure and emit copious UV radiation, which would then set a higher ionisation degree, closer to unity\cite{hosokawa2009}. This stage might not have been reached yet by G35.2N and at current high accretion rates, dust would quench the full ionisation of the disc surface, therefore keeping its wind mostly neutral. Another possible scenario to explain the low $\chi_\mathrm{e}$ is that if the central engine is accreting at high rates, then the protostar is still swollen and too cool to photo-ionise much. Both scenarios are in agreement with the fact that the ionised mass-loss rate on source is very similar to the ionised mass-loss rate on the studied knots. This implies that the $\chi_\mathrm{e}$ should not be much higher very close to the star or that the total mass-loss rate has dramatically changed (1 order of magnitude) in the past few 100 years with respect to the knots K3 and K4. This reasoning also strengths the idea that the UC/HC\,\ion{H}{ii} region, if present, must be very small and confined within $100$\,au\cite{beltran2016G35N}. Anyway, in extreme cases, some jets from HMYSOs are even observed to emerge from hypercompact \ion{H}{ii} regions\cite{guzman2016}. Our results do indicate that ionising photons from the protostar are confined to a small region, which constrains models of both evolutionary state and feedback of this source. Thus, application of these methods to a larger sample can help probe these processes more generally during massive star formation.

Finally the wider implications of our findings should be mentioned. The high collimation, velocities and momentum injection rate observed in HMYSO jets suggest that they must be (magneto-)centrifugally launched from very deep in the star's gravitational well. Coupling of magnetic fields to the gas requires a sufficient level of ionisation: the values we observe are high enough for efficient coupling\cite{pudritz2007} and are an important constraint for theoretical models of these jets and outflows\cite{staff2018,kolligan2018}, which are likely to be the dominant feedback mechanism even in massive protostellar cores\cite{tanaka2017}. The measurements that we present here are relevant to conditions at locations that are away from the disc and jet launching region. These results are important for the dynamics of outflows at distances of few $1000$ to few $10\,000$\,au from the protostar, where processes of jet collimation and outflow-core interaction are likely to be occurring. However, observations at higher spatial and spectral resolution would be required to probe the immediate jet environment close to the young star, where the jet launching mechanism could be revealed. Simulations at high resolution have been able to differentiate between a magneto-centrifugally launched highly collimated jet and a slow wide angle magnetic-pressure driven tower flow\cite{kolligan2018}, but our observations cannot discriminate between them. If we accept that indeed jets from HMYSOs are magneto-centrifugally driven, then it should be noted that earlier magnetohydrodynamical (MHD) simulations estimate lower jet velocities than those observed here, at small scales \cite{commerson2011,hennebelle2011,seifried2012} and at large scales\cite{peters2011}. However, more recent MHD simulations obtained velocities of several $100\mathrm{\,km\,s^{-1}}$ for their magneto-centrifugal jets\cite{kolligan2018}, consistent with our results. In terms of mass ejection and momentum rates, MHD simulations reproduce a wide range of values depending on the magnetic and rotational energies\cite{seifried2012,kolligan2018}, being difficult to discriminate between the different simulations. Ref.~\cite{staff2018} provides dynamical estimates remarkably close to our observations, in particular, the authors obtain mass-loss and momentum rates of the order of $\sim10^{-5}\,M_\odot\mathrm{\,yr^{-1}}$ and $\sim10^{-2}\,M_\odot\mathrm{\,yr^{-1}\,km\,s^{-1}}$, respectively, for an $8\,M_\odot$ protostar. Nevertheless, simulated outflow properties are highly dependent on resolution (see ref.~\cite{kolligan2018} for an overview of convergence aspects and numerical resolution problems in previous simulation works).


\begin{methods}

\noindent\textbf{NIR VLT/ISAAC imaging and spectroscopy}\\
NIR imaging was obtained on 2013 July 7 in the $K$ and H$_2$ bands with the ISAAC instrument on VLT (ESO, Chile). This observational strategy allows us to subtract the continuum emission of the nebulosity to obtain the H$_2$ emission line of the system (see Fig.~\ref{fig:intro} panel b, white contours). We correct the narrow band image (H$_2$) by subtracting the continuum in the broad band image ($K$), after first having normalised using the counts of the field stars. Additionally, high-resolution long-slit spectra were obtained with VLT/ISAAC on 2013 September 14 in the $H$ and $K$ bands. The $0.3''\times120''$ slit was positioned at the brightest point in the NIR with a position angle of $-14.47^\circ{}$ (see Fig.~\ref{fig:PVD} panel a, green lines). The spatial sampling was 146\,mas\,pixel$^{-1}$ and the spectral resolution was $\mathcal{R}\sim10\,000$ and $8\,900$, corresponding to a velocity resolution of $30-35\mathrm{\,km\,s^{-1}}$, for the $H$ and $K$ bands, respectively. Total integration time was 1440\,s for each band. The spatial resolution was seeing-limited at $\sim0.8-1''$. The standard $ABBA$ nodding technique was applied and the data were reduced in the standard way using the Image Reduction and Analysis Facility software (IRAF, \url{http://ast.noao.edu/data/software}). Observations were conducted under programs ID 290.C-5060(A) and 290.C-5060(B), PI: A. Caratti o Garatti.

\noindent\textbf{NIR HST/WFC3 imaging}\\
HST data were taken on 2016 March 7, proposal identifier: 14465, PI: J. Tan. Imaging was obtained in the $J$ (F110W), $H$ (F160W) wide filters, and in the [\ion{Fe}{ii}] (F164N) narrow filter using the WFC3. Total exposure time was 400, 350, and 900\,s for the $J$, $H$, and [\ion{Fe}{ii}] filters, respectively. We retrieved the pipeline-calibrated images from the HST archive. Figure~\ref{fig:intro} panel a is a three colour composite image, red is [\ion{Fe}{ii}], green is $H$, and blue is $J$. The image [\ion{Fe}{ii}]-$H$ shown in panel b of Figure~\ref{fig:intro} has been generated in a similar way to the H$_2$-$K$ image explained in the previous section.

\noindent\textbf{Radio VLA imaging}\\
VLA observations were carried out on 2012 December 29, at 5.8 GHz, with the VLA in the A-configuration ($\theta_\mathrm{beam}\sim0.3^{\prime\prime}$) and utilising a total bandwidth of $2\,\mathrm{GHz}$. Flux/bandpass and gain calibrators used were 3C286 and J1824+1044, respectively. Data reduction was performed in the usual way using the CASA software package (\url{https://casa.nrao.edu/}) together with the CASA pipeline (version 4.7.2). After several rounds of phase-only self-calibration, an RMS noise level in the image of $\sigma=6.4\,\mu\mathrm{Jy}\,\mathrm{beam}^{-1}$ was achieved. Flux densities can be found in Table~\ref{tab:radio_properties}, which are taken from \cite{purser2017thesis}. Observations were conducted under program ID 12B-140, PI: M. Hoare.

\noindent\textbf{Optical depth}\\
In order to investigate the optical depth of the knots under study, we calculate their spectral indexes. To do so, we used the radio peak intensities in the C band ($\nu_1=5.8$\,GHz) in A-configuration\cite{purser2017thesis} and K band ($\nu_2=23.0$\,GHz) in B-configuration\cite{beltran2016G35N}. This combination of bands and array configurations allows us to compare similar emission (see Table~\ref{tab:radio_properties} for radio properties of the knots).

\begin{table}[ht]
	\centering
	\caption{Radio fluxes and spectral indexes in the various knots. The source of uncertainties is coming from measurement uncertainties.}
    \vspace{5 pt}
	\label{tab:radio_properties}
	\scalebox{0.8}{
	\begin{tabular}{lcccccc}
	\hline\hline       
	\noalign{\smallskip}
		Knot & $I^\mathrm{peak}_{5.8}$ & $S_{5.8}$ & $^*I^\mathrm{peak}_{23.0}$ & $^*S_{23.0}$ & $\alpha$\\
        & (mJy/beam) & (mJy) & (mJy/beam) & (mJy) & \\
	\noalign{\smallskip}
	\hline              
	\noalign{\smallskip}
	    K1 & $0.25\pm0.01$ & $0.4\pm0.03$ & $0.15\pm0.03$ & $0.15\pm0.03$ & $-0.35\pm0.15$\\
	    K2 & $<0.19$ & $<0.19$ & $<0.10$ & $<0.10$ & $\cdots$\\
		K3 & $0.42\pm0.01$ & $1.01\pm0.03$ & $0.28\pm0.03$ & $0.60\pm0.05$ & $-0.29\pm0.09$\\
		K4 & $0.61\pm0.01$ & $2.13\pm0.14$ & $0.40\pm0.03$ & $1.10\pm0.07$ & $-0.29\pm0.09$\\
	\noalign{\smallskip}
	\hline
	\end{tabular}}
	{\small $^*$Fluxes taken from ref. \cite{beltran2016G35N}}
\end{table}

Additionally, we calculate the optical depth of the knots. For C band we obtain $\tau^{K1}_{5.8\,\mathrm{GHz}}=(2.1\pm0.4)\times10^{-2}$ $\tau^{K3}_{5.8\,\mathrm{GHz}}=(2.6\pm0.2)\times10^{-2}$ and $\tau^{K4}_{5.8\,\mathrm{GHz}}=(2.5\pm0.3)\times10^{-2}$ for knots K1, K3, and K4, respectively, showing that the emission in these knots is optically thin.

\noindent\textbf{Electron density}\\
The electron density can be estimated through the ratio of the [\ion{Fe}{ii}] lines using a non-local thermodynamic equilibrium model (NLTE) that considers the 16 fine-structure levels\cite{nisini2002,takami2006}. In our $H$ band spectroscopic ISAAC observations we detect three [\ion{Fe}{ii}] lines, the so-called $a^4D_{7/2}-a^4F_{9/2}$ at $1.644\,\mu$m, $a^4D_{1/2}-a^4F_{5/2}$ at $1.664\,\mu$m, and $a^4D_{5/2}-a^4F_{7/2}$ at $1.677\,\mu$m (see Table~\ref{tab:NIR_properties} for the fluxes of the various knots). The ratio 1.664/1.644 provides a good estimate of the electron density whereas the ratio between 1.644/1.677 provides a less accurate estimate because this ratio is less sensitive to the electron density at values higher than few $10^4\mathrm{\,cm^{-3}}$.

\begin{table}[ht]
	\centering
	\caption{[\ion{Fe}{ii}] lines observed in the various knots. The source of uncertainties is coming from measurement uncertainties.}
    \vspace{5 pt}
	\label{tab:NIR_properties}
	\scalebox{0.8}{
	\begin{tabular}{lcccc}
	\hline\hline       
	\noalign{\smallskip}
		Knot & $F_{1.644}$ & $F_{1.664}$ & $F_{1.677}$\\
        & & $(\times10^{-14}\mathrm{\,erg\,s^{-1}\,cm^{-2}})$\\
	\noalign{\smallskip}
	\hline              
	\noalign{\smallskip}
	    K1$^*$ & $1.03\pm0.10$ & $\cdots$ & $\cdots$\\
	    K2 & $0.18\pm0.01$ & $<0.02$ & $<0.02$\\
		K3 & $1.46\pm0.04$ & $0.17\pm0.03$ & $0.33\pm0.05$\\
		K4 & $2.90\pm0.10$ & $0.39\pm0.10$ & $0.79\pm0.14$\\
	\noalign{\smallskip}
	\hline
	\end{tabular}}
	\\$^*$ {\small The flux was taken from the HST image because our ISAAC slit did not encompass knot K1.}
\end{table}

The electron density can also be computed using the radio properties of the knots following Equation (7) of ref. \cite{mezger1967}:

\begin{align}
\left(\frac{n_e}{\mathrm{cm^{-3}}}\right) =\, & u_1 \cdot6.884\times10^2\cdot\left(\frac{T_e}{10^4\,K}\right)^{0.175}\cdot\left(\frac{S_\mathrm{5\,GHz}}{\mathrm{Jy}}\right)^{0.5}\cdot \nonumber \\ & \cdot\left(\frac{D}{\mathrm{kpc}}\right)^{-0.5}\cdot\left(\frac{\theta_G}{\mathrm{arcmin}}\right)^{-1.5}
\label{eq:mezger_ne}
\end{align}

\noindent where $u_1=0.857$ is the density model II (cylinder) conversion factor for computing electron density, $T_e$ is the electron temperature, $S_\mathrm{5\,GHz}$ is the flux density at 5\,GHz, $D$ is the distance to the source, and $\theta_G=(\theta_\mathrm{min}\cdot\theta_\mathrm{maj})^{0.5}$ is the deconvolved Gaussian width of the knot. This equation is valid if the emission is optically thin, which is the case of the knots under consideration (see previous section). Using Equation \ref{eq:mezger_ne} and considering $T_e=10^4$\,K, $D=2.2$\,kpc, the flux density (from Table~\ref{tab:radio_properties}, assuming $\alpha=-0.1$), and the size of each knot (from Table~\ref{tab:mass-loss_radio}); values of the order of $10^4\mathrm{\,cm^{-3}}$ were obtained for the various knots (see Table~\ref{tab:electron_density}, Column 2). As we obtain spectral indices of $\sim-0.3$ in our knots, there might be some mixture between thermal and non-thermal emission. To remove any non-thermal contribution, we assume $\alpha=-0.1$ and extrapolate the flux density from the 23\,GHz data. Anyhow it is worth noting that the flux dependence is to the power of 0.5 and, therefore, there is no significant change (less than a factor of 1.5) in the electron number estimates.

It is worth mentioning that in the case of knot K3, where we are able to detect the [\ion{Fe}{ii}] lines with enough S/N ratio (see Table~\ref{tab:NIR_properties}) to estimate a reliable electron density, the estimates for the electron density from the NIR and the radio coincide within the errors (see Table~\ref{tab:electron_density}). The same applies for knot K4, but here the uncertainties are much larger because of the low S/N ratio (see Table~\ref{tab:NIR_properties}). Therefore, the two independent methods provide us with the same result. In the case of knot K4, however, the S/N ratio in the [\ion{Fe}{ii}] lines is quite low to estimate a reliable electron density from the NIR as is evidenced by the larger errors in this knot. We could still calculate a more precise electron density with smaller errors from the radio regime. For knot K1 no information about the electron density could be retrieved from the NIR because our slit did not encompass this knot and only radio emission could be used to calculate the electron density. Finally, for knot K2, only upper limits for the electron density could be given because the fluxes of the lines $1.664\mathrm{\,\mu m}$ and $1.677\mathrm{\,\mu m}$ are below $3\sigma$ (see Table~\ref{tab:NIR_properties}). For consistency, we then adopt the electron density derived from the radio for the four knots.

\begin{table}[ht]
	\centering
	\caption{Electron density $(n_\mathrm{e})$ for the various knots. The source of uncertainties in Column 2 is coming from the error propagation calculated from Equation~\ref{eq:mezger_ne}, whereas in Columns 3 and 4 from the error propagation of the ratio of the lines together with the NLTE model.}
    \vspace{5 pt}
	\label{tab:electron_density}
	\scalebox{0.8}{
	\begin{tabular}{lcccc}
	\hline\hline       
	\noalign{\smallskip}
		Knot & $n^\mathrm{radio}_\mathrm{e}$ & $n^\mathrm{1.664/1.644}_\mathrm{e}$ & $n^\mathrm{1.644/1.677}_\mathrm{e}$\\
        & $(\times10^{4}\mathrm{\,cm^{-3}})$ & $(\times10^{4}\mathrm{\,cm^{-3}})$ & $(\times10^{4}\mathrm{\,cm^{-3}})$\\
	\noalign{\smallskip}
	\hline              
	\noalign{\smallskip}
	    K1 & $1.8\pm0.1$ & $\cdots$ & $\cdots$\\
	    K2 & $<0.35-0.6$ & $<0.65$ & $<1.0$\\
		K3 & $2.1\pm0.1$ & $2.3\pm0.1$ & $3.0\pm2.0$\\
		K4 & $1.6\pm0.1$ & $3.0\pm2.0$ & $6.0\pm5.0$\\
	\noalign{\smallskip}
	\hline
	\end{tabular}}
\end{table}

\noindent\textbf{Ionisation fraction}\\
In the following the derivation of the ionisation fraction is outlined (the justification for our assumptions is given below). The ionisation fraction is defined as Equation~\ref{eq:ionisation_fraction}:

\begin{equation}
    \chi_e = n_e/n_\mathrm{tot}
    \label{eq:ionisation_fraction}
\end{equation}

\noindent where $n_\mathrm{tot}$ is the total number density and $n_\mathrm{e}$ is the electron number density. On the one hand, the number electron density is calculated following the previous section. On the other hand, the total number density can be derived following the next steps. Firstly, the product $n_\mathrm{tot}V$, where $V$ is the volume emitting region, can be expressed as the ratio between the observed luminosity of the line and the emissivity per particle for the line calculated theoretically\cite{podio2006}. That is, $n_\mathrm{tot}V=L_\mathrm{line}/\epsilon_\mathrm{line}=L_\mathrm{line}\left(h \nu A_i f_i \frac{Fe^+}{Fe}\left[\frac{Fe}{H}\right]\right)^{-1}$ where $A_i$ and $f_i$ are the radiative rates and fractional population of the upper level of the considered transition and $\frac{Fe^+}{Fe}$ is the ionisation fraction of the iron having a total abundance with respect to hydrogen of $\left[\frac{Fe}{H}\right]$\cite{nisini2005,garcia-lopez2008}. Here, we have used the [\ion{Fe}{ii}] line at $1.644\,\mu$m. We have assumed that all iron is ionised and solar abundance of $2.8\times10^{-5}$ under the hypothesis of no dust depletion\cite{asplund2005}. Secondly, both the radius and the length of the knots are resolved in our HST and VLT/ISAAC, therefore we can calculate the precise value of the volume. Considering that the volume of the knot is a cylinder with radius $r_\mathrm{jet}=D\cdot\tan(\theta_\mathrm{min}/2)\simeq D\cdot\theta_\mathrm{min}/2$, where $\theta_\mathrm{min}$ is the deconvolved diameter of the knot (see Fig.~\ref{fig:jet_width} and Table~\ref{tab:mass-loss_FeII} Column 6), and with length $l_\perp = D\cdot\tan(\theta_\mathrm{maj})\simeq D\cdot \theta_\mathrm{maj}$ where $\theta_\mathrm{maj}$ is the deconvolved length of the knot (see Table~\ref{tab:mass-loss_FeII} Column 5), we can write $V=\pi r_\mathrm{jet}^2 l_\perp$, being $D$ the distance to the source. Finally, solving for $n_\mathrm{tot}$, we can derive the total number density of the region given by Equation~\ref{eq:n_tot}:

\begin{equation}
n_\mathrm{tot}=L_\mathrm{line}/\epsilon_\mathrm{line}/V
\label{eq:n_tot}
\end{equation}

From a theoretical point of view, iron is easily ionised ($E_\mathrm{ion} = 7.9$\,eV) in mild J-shocks ($v_\mathrm{shock}\sim25-50\mathrm{\,km\, s^{-1}}$, see e.g., ref.\,\cite{mcCoey2004,koo2016}); in the case of G35.2N we are dealing with much faster shocks ($v_\mathrm{shock}\geq500\mathrm{\,km\, s^{-1}}$). The assumption that all iron is ionised is also supported by our observations where both [\ion{Fe}{ii}] and ionised hydrogen ($E_\mathrm{ion} = 13.6$\,eV) emissions are co-spatial. Moreover, no [\ion{Fe}{i}] emission has ever been detected in protostellar jets driven by HMYSOs. Indeed, [\ion{Fe}{i}] has seldom been observed in protostellar jets driven by low-mass YSOs. For example, ref.\,\cite{lahuis2010} observed in the mid-infrared (less affected by the visual extinction) 61 low-mass embedded sources. This study reports [\ion{Fe}{ii}] emission (at $17.94$ and $25.99\mathrm{\,\mu m}$) in 13 out of 61 sources, whereas [\ion{Fe}{i}] emission (at $24.04 \mathrm{\,\mu m}$) is only detected in one source where no [\ion{Fe}{ii}] is observed. The assumption of no dust depletion is sustained by the fact that the faster the shock velocity, the higher the fraction of the dust-forming elements transferred into the gas-phase\cite{jones2000}. Indeed, shocks with velocities $\geq400\mathrm{\,km\,s^{-1}}$ completely destroy the dust\cite{draine2003} and release the Fe grains into gas-phase. We certainly measure jet velocities $>500\mathrm{\,km\,s^{-1}}$.

On the other hand, it should be noted that ref.\cite{podio2006} found that Fe depletion can reach up to $\sim90\%$ for the HH\,34 jet, driven by a low-mass YSO. However, the jet we are analysing here is driven by a HMYSO, and therefore more energetic shocks are measured. Indeed, the line profile of less energetic jets as HH\,34 is very narrow up to large distances from the source (see e.g. Fig.\,3 of ref.\cite{garcia-lopez2008}). In the case of our Figure~\ref{fig:PVD} panel b, it is clear that the line profile is different, resembling that of a bow-shock. In addition, later studies on HH\,34 and other low-mass jets (see ref.\cite{podio2009} and ref.\cite{nisini2016}) show that [\ion{Fe}{ii}] is mostly detected in the low-velocity component close to the source whereas the high-velocity component show very low or no dust depletion. In any case, if the scenario of high depletion were in place in our study, this would imply that $n_\mathrm{tot}$ is underestimated by an order of magnitude and, therefore, that the ionisation fraction is also overestimated by an order of magnitude.

For knots K3 and K4 we are able to provide a good estimate of the ionisation fraction. However, for knot K2 only an upper limit for the electron density could be obtained and therefore an upper limit for the ionisation fraction is given. Finally, for knot K1 the uncertainty is quite large due to that the knot geometry is not very clear in our NIR images and we have considered larger errors.

\noindent\textbf{Mass-loss and momentum rate}\\
The mass-loss rate was computed following the equation:

\begin{equation}
\dot{M}_\mathrm{ejec}=\frac{M v_\perp}{l_\perp}
\label{eq:mass_loss}
\end{equation}

\noindent where $M$ is the mass of the knot, $v_\perp$ the tangential velocity of the knot, and $l_\perp$ the length of the knot in the plane of the sky. The mass of the knot is not a direct measurement, but it can be written as $M=\mu m_H n_\mathrm{tot}V$ where $\mu=1.24$ is the mean atomic weight, $m_H = 1.67\times10^{-27}$\,kg is the mass of the hydrogen, $n_\mathrm{tot}$ is the total density of the hydrogen (both neutral and ionised) of the flow, and $V$ is the volume of the emitting region (assumed to be a cylinder). We derived above that the product $n_\mathrm{tot}V$ can be expressed as the ratio between the line luminosity and the line emissivity of the [\ion{Fe}{ii}] line at 1.644$\mu$m. Therefore, the Equation \ref{eq:mass_loss} can be written as Equation~\ref{eq:mass_loss_feii}:

\begin{equation}
\dot{M}_\mathrm{ejec}=\mu m_H L_\mathrm{[\ion{Fe}{ii}]1.644} \left(h \nu A_i f_i \frac{Fe^+}{Fe}\left[\frac{Fe}{H}\right]\right)^{-1} v_\perp \,\frac{1}{l_\perp}
\label{eq:mass_loss_feii}
\end{equation}

\noindent The luminosity of the line is expressed as $L=4\pi D^2 F$, where $D$ is the distance to the source from Earth and $F$ the dereddened flux using the Rieke \& Lebofsky extinction law\cite{rieke1985} considering $A_V = 25$\,mag\cite{fuller2001}. The input parameters values and the mass-loss rates can be found in Table~\ref{tab:mass-loss_FeII}.

\begin{table*}[ht]
	\centering
	\caption{Input parameters and mass-loss rates for the [\ion{Fe}{ii}] 1.644$\mu$m line. The source of uncertainties in Columns 4, 5, and 6 is coming from measurement uncertainties, whereas in Columns 2, 3, 7, 8, and 9 comes from performing the error propagation in the specific equations.}
    \vspace{5 pt}
	\label{tab:mass-loss_FeII}
	\scalebox{0.85}{
	\begin{tabular}{lccccccccc}
	\hline\hline       
	\noalign{\smallskip}
		Knot & $L_\mathrm{[\ion{Fe}{ii}]}$ & $n_\mathrm{tot}$ & $v_\perp$ & $\theta_\mathrm{maj}$ & $\theta_\mathrm{min}$ & $M$ & $\dot{M}_\mathrm{ejec}$ & $\dot{P}$\\
        & $(\times10^{32}\mathrm{\,erg\,s^{-1}})$ & $(\times 10^5\mathrm{\,cm^{-3}})$ & $(\mathrm{km\,s^{-1}})$ & (arcsec) & (arcsec) & ($\times10^{-3}\,M_\odot$) & $(\times10^{-5}\,M_\odot\mathrm{\,yr^{-1}})$ & $(\times10^{-2}\,M_\odot\mathrm{\,yr^{-1}\,km\,s^{-1}})$\\
	\noalign{\smallskip}
	\hline              
	\noalign{\smallskip}
	    K1 & $3.43\pm0.69$ & $1.5\pm0.7$ & $150-300^*$ & $1.5\pm0.2$ & $0.5\pm0.2$ & $1.6\pm0.3$ & $1.6-3.2$ & $0.2-0.9$\\
	    K2 & $0.58\pm0.12$ & $0.36\pm0.1$ & $140\pm10$ & $1.5\pm0.1$ & $0.53\pm0.01$ & $0.3\pm0.1$ & $\cdots$ & $\cdots$\\
		K3 & $4.86\pm0.13$ & $3.0\pm0.2$ & $355\pm10$ & $1.7\pm0.1$ & $0.40\pm0.01$ & $2.4\pm0.1$ & $4.7\pm0.3$ & $1.8\pm0.1$\\
		K4 & $9.12\pm0.33$ & $3.1\pm0.2$ & $205\pm10$ & $1.8\pm0.1$ & $0.52\pm0.01$ & $4.5\pm0.2$ & $4.9\pm0.3$ & $1.1\pm0.1$\\
	\noalign{\smallskip}
	\hline
	\end{tabular}}
	\\$^*$ Assuming a tangential velocity of $150-300\mathrm{\,km\,s^{-1}}$ from the proper motions of ref.\cite{beltran2016G35N}.	
\end{table*}

The mass-loss rate can also be calculated using the radio emission properties of the various knots. In this case, the mass of the knot is calculated as $M=V\cdot n_e\cdot\mu/\chi_e$, where $V$ is the volume of the knot (considered to be a cylinder as in the case of the NIR), $n_\mathrm{e}$ is the number electron density calculated in the previous section, $\mu$ is average atomic weight, and $\chi_e$ is the ionisation fraction (see previous section). The ejection time, which is the period of time that the jet lobe was ejected over, is equal to $\tau_\mathrm{ejec}=D\cdot\tan(\theta_\mathrm{maj})/v_\perp$. See Table~\ref{tab:mass-loss_radio} Columns 2 and 3 for the deconvolved dimensions of the emission lobes. Finally, the mass-loss rate is then given by $\dot{M}_\mathrm{ejec}=M/\tau_\mathrm{ejec}$, see Table~\ref{tab:mass-loss_radio} for input parameters and mass-loss rates estimates (Column 5 for ionised 
mass-loss rate where 
$\chi_\mathrm{e}$ was simply considered equal to $1$, Column 6 for total mass-loss rate considering the ionisation fraction calculated above).

The mass-loss rates for knots K1 and K2 are quite uncertain. In the case of K1, no reliable velocity could be measure; and in the case of K2, the electron density could not be retrieved, just an upper limit.

Finally, the momentum rate (or thrust) is given by $\dot{P}=\dot{M}_\mathrm{ejec}v_\mathrm{tot}$ (see Column 9 of Table~\ref{tab:mass-loss_FeII} for the momentum rate of NIR jet, and Column 7 of Table~\ref{tab:mass-loss_radio} for the momentum rate of the ionised radio jet, $\dot{P}_\mathrm{ionised}$).

\begin{table*}[ht]
	\centering
	\caption{Input parameters and mass-loss rates for the Radio C-band (6 cm). The source of uncertainties in Columns 2 and 3 is coming from measurement uncertainties and in Columns 4, 5, and 6 performing the error propagation in the specific equations.}
    \vspace{5 pt}
	\label{tab:mass-loss_radio}
	\scalebox{0.9}{
	\begin{tabular}{lcccccc}
	\hline\hline       
	\noalign{\smallskip}
		Knot & $\theta_\mathrm{maj}$ & $\theta_\mathrm{min}$ & $M_\mathrm{ionised}$ & $\dot{M}_\mathrm{ionised}$ & $\dot{M}_\mathrm{ejec}$ & $\dot{P}_\mathrm{ionised}$\\
        &  (arcsec) & (arcsec) & ($\times10^{-4}\,M_\odot$) & $(\times10^{-6}\,M_\odot\mathrm{\,yr^{-1}})$ & $(\times10^{-5}\,M_\odot\mathrm{\,yr^{-1}})$ & $(\times10^{-4}\,M_\odot\mathrm{\,yr^{-1}\,km\,s^{-1}})$\\
	\noalign{\smallskip}
	\hline              
	\noalign{\smallskip}
	    K1$^*$ & $0.34\pm0.04$ & $0.20\pm0.03$ & $0.7\pm0.3$ & $0.3-0.6$ & $0.26-0.52$ & $0.45-1.8$\\
	    K2 & $\cdots$ & $\cdots$ & $\cdots$ & $\cdots$ & $\cdots$ & $\cdots$\\
		K3 & $0.61\pm0.03$ & $0.23\pm0.01$ & $2.0\pm0.5$ & $1.1\pm0.3$ & $1.6\pm0.4$ & $3.9\pm0.9$\\
		K4 & $0.93\pm0.07$ & $0.33\pm0.03$ & $4.8\pm1.7$ & $1.0\pm0.3$ &$2.0\pm0.7$ & $2.3\pm0.7$\\
	\noalign{\smallskip}
	\hline
	\end{tabular}}
	\\$^*$ Assuming a tangential velocity of $150-300\mathrm{\,km\,s^{-1}}$ from the proper motions of ref.\cite{beltran2016G35N}.
\end{table*}

\begin{figure*}[ht!]
\includegraphics[width=1.0\textwidth]{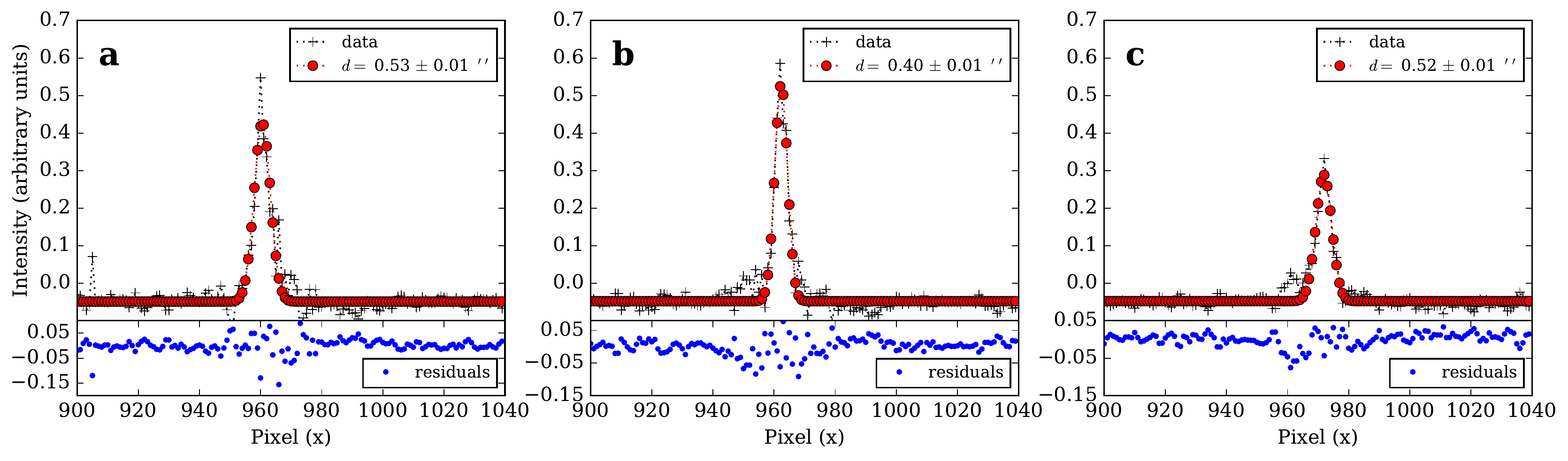}
\caption{Gaussian fitting to the jet width using the HST [\ion{Fe}{ii}] $1.644\,\mu\mathrm{m}$ continuum-subtracted image. Black data points are the data, red data points are the best Gaussian fit, and the blue dots are residuals from the fit. Best fit values for the jet width are indicated in the figure legend for the K2, K3, and K4 knots in panels a, b, and c, respectively.}
\label{fig:jet_width}
\end{figure*}

\noindent\textbf{Ionised mass-loss rate on source}\\
The ionised mass-loss rate can be calculated following the Reynolds' formulation\cite{reynolds1986} (see also ref.\cite{anglada2018}). For a canonical jet $(\alpha=0.6)$ the ionised mass-loss rate is given by Equation~\ref{eq:mass_loss_ionised}:

\begin{align}
\left(\frac{\dot{M}_\mathrm{ionised}}{10^{-6}\,M_\odot\mathrm{\,yr^{-1}}}\right) =\, & 0.139\left[\left(\frac{S_\nu}{\mathrm{mJy}}\right)\left(\frac{\nu}{10\mathrm{\,GHz}}\right)^{-0.6}\right]^{0.75}
\nonumber\\& \times \left(\frac{v_j}{200\mathrm{\,km\,s^{-1}}}\right)\left(\frac{\theta_0}{\mathrm{rad}}\right)^{0.75}(\sin i)^{-0.25}
\nonumber\\& \times \left(\frac{D}{\mathrm{kpc}}\right)\left(\frac{T}{10^4\mathrm{\,K}}\right)^{-0.075}
\label{eq:mass_loss_ionised}
\end{align}

\noindent where $S_\nu$ is the flux density, $\nu$ is the frequency, $v_j$ is the jet velocity, $\theta_0$ is the jet opening angle, $i$ is the inclination of the jet, $D$ is the distance, and $T$ the temperature. From our observations we obtain the following parameters for core B: $S_{5.8}=0.794\pm0.03$\,mJy, $\nu = 5.8$\,GHz, $v_j = 600\pm100\mathrm{\,km\,s^{-1}}$,  $\theta_0 = 52.3^\circ{}\pm4.4^\circ{}$, $D = 2.2$\,kpc, $T = 10000$\,K, the resulting ionised mass-loss rate is $1.81\pm0.33\times10^{-6}\,M_\odot\mathrm{\,yr^{-1}}$, consistent with ref.\,\cite{beltran2016G35N}. This value is very similar to those found along the knots (see Table~\ref{tab:mass-loss_radio}, Column 5). This indicates that has not been a great variation in the ionised mass-loss rate between the knots located up to $\sim18\,000$\,au and the jet very close to the massive protostar.

\newpage
\noindent\textbf{Data availability}\\
All data used in this study are public and can be accessed through the different data archives of the various instruments using the Program ID and/or PI name given in the section above. ESO archive: \url{http://archive.eso.org/eso/eso_archive_main.html}, HST archive: \url{http://hst.esac.esa.int/ehst/#home}, VLA archive: \url{https://archive.nrao.edu/archive/advquery.jsp}. Upon reasonable request the authors will provide all data supporting this study.

\noindent\textbf{Code availability}\\
Upon reasonable request the authors will provide all code supporting this study.

\end{methods}


\newpage
\noindent\textbf{References}
\bibliography{phd_bibliography}{}
\bibliographystyle{nature}


 \noindent\textbf{Acknowledgements}\\
 R.F. acknowledges support from Science Foundation Ireland (grant 13/ERC/12907). A.C.G., S.J.D.P, and T.P.R. have received funding from the European Research Council (ERC) under the European Union's Horizon 2020 research and innovation programme (grant agreement No.\ 743029). R.G.L has received funding from the European Union's Horizon 2020 research and innovation programme under the Marie Sk\l{}odowska-Curie Grant (agreement No.\ 706320). J.C.T. acknowledges support from NSF grant AST 1212089 and ERC Advanced Grant 78882 (MSTAR). This research is based on observations collected at the VLT (ESO Paranal, Chile) with programme IDs 290.C-5060(A)  and  290.C-5060(B). This research is based on observations made with the NASA/ESA Hubble Space Telescope with proposal identifier: 14465. The Karl G. Jansky Very Large Array (VLA) observations presented here were part of National Radio Astronomy Observatory program 12B-140. The National Radio Astronomy Observatory is a facility of the National Science Foundation operated under cooperative agreement by Associated Universities, Inc.
 
Throughout this work, we made use of astropy \url{http://www.astropy.org}, a community-developed core PYTHON package for astronomy (version 3.0.3), and uncertainties, a PYTHON package for calculations with uncertainties (version 3.0.2) developed by Eric O. Lebigot \url{http://pythonhosted.org/uncertainties/}, for plotting and error propagation purposes, respectively.
 
\noindent\textbf{Author contributions}\\
R.F, A.C.G., and T.P.R. wrote the initial manuscript. R.F, A.C.G., R.G.L, S.J.D.P., and D.C. worked on the data reduction and analysis. A.C.G., J.C.T. and M.H. are the PIs of the ESO, HST, and the VLA proposals, respectively. R.G.L, A.S., and B.S, are coauthors of the proposals. All coauthors commented on the manuscript.

\noindent\textbf{Additional information}\\
Reprints and permissions information is available at www.nature.com/reprints

\noindent\textbf{Correspondence}\\
Correspondence and requests for materials should be addressed to R.F.~(email: fedriani@cp.dias.ie).

\noindent\textbf{Competing Interests}\\
The authors declare that they have no competing interests.

\end{document}